# Local operation on identical particles systems.

**O Rendón[1], and E Medina[2].**

[1]Departamento de Física – FACYT, Universidad de Carabobo, Valencia, Edo. Carabobo, Venezuela.

[2]Instituto Venezolano de Investigaciones Científicas, IVIC, Apartado 21827, Caracas 1020 A, Venezuela.

E-mail: orendon@uc.edu.ve

**Abstract**. We describe identical particles through their extrinsic physical properties and operationally with an operator of selective measure $M_m$. The operator $M_m$ is formed through non-symmetrized tensor product of one-particle measurement operators, so that it does not commute with the permutation operator P. This operator of selective measure $M_m$ is a local operation (LO) if it acts on physical systems of distinguishable particles, but when $M_m$ acts on the Hilbert sub-space of a system with a constant number of indistinguishable particles this can generate entanglement in the system. We will call entanglement by measurement (EbM), this way of producing entanglement. In this framework, we show entangle production examples for systems of two fermions (or bosons) when the operator $M_m$ has two-particle events that are not mutually exclusive.

## 1. Introduction

In the basic texts of the quantum mechanics it is very well-known that the identical particles are indistinguishable, so that an observable that distinguishes the intrinsic labels of each particle cannot exist when the wave functions of the individual particles are overlapped, [1]. The indistinguishable approach consists in not to privileging the presence of each particle in one of the quantum states of one-particle, then it is fundamentally ignored which particle is in each one-particle quantum state. In consequence, for indistinguishable particles the wave function is antisymmetrized (fermions) and symmetrized (bosons) and therefore the definition of entangled states as in the case of distinguishable particles does not apply, [2].

Bi-partite entanglement has been studied with different focuses, see [3]: An approach based on the structure of the tensorial product of composite systems ([4], [5] and [6]), one based on the occupation-number representation ([7] and [8]) and another on the mixture of both [9]. In spite of these efforts the indistinguishable particles issue is hard to kill: [3] Horodecki quote: "However, it seems that there is still controversy concerning the meaning of entanglement for indistinguishable particles, [9]"

In the ideal Fermi system, spin entanglement between two distant particles in the ground state, at T= 0 K, has been studied by Vedral ([10] and [11]). He concludes:

- No correlation for spin electrons with different momentum.
- Correlation for spin electrons located in different regions of space, $V \cap V´ = \emptyset$. This entanglement depends on the dimensionality and decrease with the distance among disjoint regions.

We develop the concept of "local operation" for two identical particles with possibility of being distinguished for their extrinsic properties following the ideas of: 1) Wiseman and Vaccaro [9] to project a state of N identical particles $|\psi\rangle$ into two distinct well-defined subspaces, and 2) Herbut [12] on the possibility of using both indistinguishable and distinguishable approach to describe a system of two identical particles that are well separated, e.g. $V \cap V´ = \emptyset$ or distinct spin orientations along the z axis (assuming these properties don´t change during the time of interest).

Here, we use the distinguishable particle framework and in future studies [13] we show a more comprehensive approach, considering the approach of distinguishable particles as indistinguishable and within a theoretical development based on quantum information concepts [14].

In the following we present the conditions that a "local operation" should have so that it doesn't change the entanglement of a system of two identical particles, then two examples are shown where the entanglement generation for non local operation is evident.

## 2. "Local Operation" for two identical particles

We define the local operation (LO) for a system of two identical particles, within of the distinct-particle approach. This formalism is valid for fermions and bosons. In the simple version to distinguish two identical particles, we follow to Herbut [12]:

"Actually, we want to generalize the well-know idea that if two particles can be distinguished either because they are known to be localized in well-separated spatial regions, e.g., V and V´, or because they have, e.g., distinct spin orientations along the z axis (and these properties not change during the time of interest). Then we can treat the identical particles as if they were distinct."

Let a distinguishable two-particles Hilbert space $H^{(2)}=H^{(1)}\otimes H^{(1)}$, consisting of one-particle Hilbert space factorable in $H^{(1)}=H´\otimes H´´\otimes...$, be defined for example: spatial versus spin component along the z axis ($H^{(1)}=H_r \otimes H_{Sz}$) or momentum versus spin component along the x-axis ($H^{(1)}=H_k \otimes H_{Sx}$).

The operator of selective measure $M_m$ distinguishes particles and has its domain and codomain in $H^{(2)}$, is a local operation (LO) if the following two conditions are satisfied:

a) $M_m$ is a tensor product of one-particle measure selective operator

$$M_m = M_{m1} \otimes M_{m2}, \quad (2.1)$$

generally with arbitrary domain and codomain in one-particle Hilbert space, $H^{(1)}$.

$$M_{m1}: H´´ \to H´´ \quad (2.2)$$

$$M_{m2}: H´ \to H´ \quad (2.3)$$

b) For any state of distinguishable two-particles, $|\psi\rangle \in H^{(2)}$, it has an orthogonal sum

$$|\psi\rangle = \frac{1}{\sqrt{2}}\left[|\psi\rangle_{fermions} + |\psi\rangle_{bosons}\right]. \quad (2.4)$$

The operator $M_m$ that distinguishes identical particles (fermions or bosons) does not generate entanglement by measurement (EbM) if $M_m|\psi\rangle = 0$ or $M_m P|\psi\rangle = 0$ but not both; where P is the permutation operator.

Any operator that acts on a pair of identical particles and satisfies the above two conditions is a local operation (LO). The operator of selective measure $M_m$ is not symmetrized, see equation (2.1), then it doesn't commute with the permutation operator P.

The proof of the above proposition is simple; both the Hilbert subspace associated with fermions or bosons decomposes in the same way

$$|\psi\rangle_{fermions} = \frac{1}{\sqrt{2}}[1-P]|\psi\rangle \tag{2.5}$$

$$|\psi\rangle_{bosons} = \frac{1}{\sqrt{2}}[1+P]|\psi\rangle \tag{2.6}$$

where **1** is the identity operator and P is the permutation operator. The ket $|\psi\rangle$ is the quantum state that characterizes two distinguishable particles, with $|\psi\rangle$ orthogonal to $P|\psi\rangle$.

We distinguish two fermions (or bosons) when measured with $M_m$

$$M_m|\psi\rangle_{fermions} = \frac{1}{\sqrt{2}}[M_m|\psi\rangle - M_m P|\psi\rangle] \tag{2.7}$$

$$M_m|\psi\rangle_{bosons} = \frac{1}{\sqrt{2}}[M_m|\psi\rangle + M_m P|\psi\rangle] \tag{2.8}$$

and two events are defined as: $M_m|\psi\rangle$ and $M_m P|\psi\rangle$. If one of the two events $M_m|\psi\rangle$ or $M_m P|\psi\rangle$ is zero, but not both; then $M_m$ distinguishes particles without increasing entanglement and is called local operation (LO). Otherwise, $M_m$ distinguish particles, and contemplates the two events $M_m|\psi\rangle$ and $M_m P|\psi\rangle$ with nonzero probability and $\langle\psi|M_m P|\psi\rangle = 0$, (That is, the kets $M_m|\psi\rangle$ and $M_m P|\psi\rangle$ are geometrically orthogonal); then the ket $|\psi\rangle_{fermions}$ (or $|\psi\rangle_{bosons}$) collapses to a entangled quantum state if the initial state is not entangled, see equation (2.7) or (2.8). We will call this procedure entanglement by measurement (EbM).

2.1. Example of ¨Local Operation"
Given the non-entangled two particle, [5].

$$|\psi\rangle_{\substack{fermions\\or\ bosons}} = \frac{1}{\sqrt{2}}\left[|k_a k_b\rangle|\uparrow\downarrow\rangle \mp |k_b k_a\rangle|\downarrow\uparrow\rangle\right] \tag{2.9}$$

and the selective measure operator

$$M_m = |k_a k_b\rangle\langle k_a k_b| \tag{2.10}$$

Here the state of two distinguishable particles that builds the states of indistinguishable particle $|\psi\rangle_{fermions}$ (or $|\psi\rangle_{bosons}$) is

$$|\psi\rangle = |k_a k_b\rangle|\uparrow\downarrow\rangle \tag{2.11}$$

and

$$P|\psi\rangle = |k_b k_a\rangle|\downarrow\uparrow\rangle \qquad (2.12)$$

with $k_a \neq k_b$.

In this projection, we have used momentum variables of each particle as labels to distinguish particles. We can see that $M_m P|\psi\rangle = 0$ and then $|\psi\rangle_{\substack{fermions\\or\ bosons}}$ collapses to:

$$M_m|\psi\rangle_{\substack{fermions\\or\ bosons}} = \frac{1}{\sqrt{2}}|k_a k_b\rangle|\uparrow\downarrow\rangle \qquad (2.13)$$

and this selective measure does not increase entanglement in the initial state, see equation (2.9) and (2.13).

The stat $|\psi\rangle_{\substack{fermions\\or\ bosons}}$, (2.9), describes a particle with momentum $k_a$ and spin up, and the other with momentum $k_b$ and spin down and $M_m|\psi\rangle_{\substack{fermions\\or\ bosons}}$, (2.13), characterizes the particle $k_a$ with spin up and the particle $k_b$ with spin down. Note that the momentum variables are used as labels. Within the distinguishable approach the identical two-particle state (2.9) can be expressed [15], as:

$$|\psi\rangle_{dist} = |\uparrow_{k_a} \downarrow_{k_b}\rangle \qquad (2.14)$$

2.2. An example with operation no-local.

Again, we begin with the state $|\psi\rangle_{\substack{fermions\\or\ bosons}}$, (2.9), but now it describes an one-dimensional system of two identical particles. We assume, for the sake of the argument, that the momentum $k_a = k$ y $k_b = -k$ have the same magnitude k, but opposite directions.

The state $|\psi\rangle_{\substack{fermions\\or\ bosons}}$, (2.9), is a momentum self-ket, then the position of each particle is no well-defined and the two events can happen with the same probability a) particles approach each other, or b) away from one another, see figure 1.

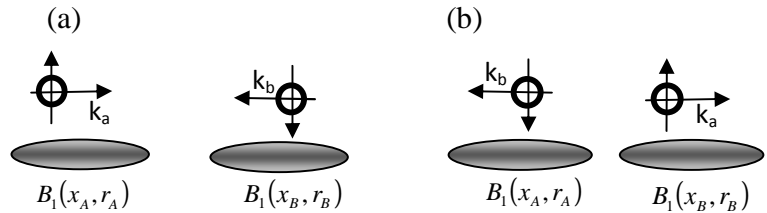

**Figure 1.** The state $|\psi\rangle_{\substack{fermions\\or\ bosons}}$, (2.9) describes two possible events
(a) particles approach each other, or (b) away from one another.

Lets take an operator of selective measure $M_m$, projector of the position self-ket.

$$M_m = \int_{B_1(x_A, r_A)} |x_1\rangle dx_1 \langle x_1| \otimes \int_{B_1(x_B, r_B)} |x_2\rangle dx_2 \langle x_2|, \tag{2.15}$$

with

$$B_1(x_A, r_A) = \left(x_A - \frac{r_A}{2}, x_A + \frac{r_A}{2}\right), \tag{2.16}$$

and

$$B_1(x_B, r_B) = \left(x_B - \frac{r_B}{2}, x_B + \frac{r_B}{2}\right). \tag{2.17}$$

In this projection, we have used position variables for each particle as labels to distinguish particles. If $r_A = r_B \approx \Delta x \ll 1$, development to the lowest order in $\Delta x$, so the ket $|\psi\rangle_{\substack{fermions \\ or\ bosons}}$ collapses to:

$$M_m |\psi\rangle_{\substack{fermions \\ or\ bosons}} = \Delta x^2 \langle x_A | k_A \rangle \langle x_B | k_B \rangle \{|x_A \uparrow\rangle |x_B \downarrow\rangle \pm e^{-i\Delta k \Delta x} |x_A \downarrow\rangle |x_B \uparrow\rangle\} \tag{2.18}$$

with $\Delta k = k_B - k_A$ and $\Delta x = x_B - x_A$.

The ket $M_m |\psi\rangle_{\substack{fermions \\ or\ bosons}}$, (2.18), describes two events and the information of the whole systems is that the particles located in different regions $B_1(x_A, r_A)$ and $B_1(x_B, r_B)$ have opposite spin component, but we do not know the orientation of the spin in each region $B_1(x_A, r_A)$ and $B_1(x_B, r_B)$, see figure 1. Here, the location of the particles has been entangled in the internal degrees of freedom (spin) and its origin is based on $M_m$ is not operation local.

**3. Conclusions**

One of the goals of quantum computation is the systematic manipulation of physical objects with quantum properties. In the context of solid state physics, a common task in quantum computing is to use electrons in a material and separate them to two different users, Alice and Bob. The assignment of a quantum system of identical particles to each user, Alice and Bob, is a process of distinguishing identical particles.

The process of associating an identical particle to each user is to give an electron to Alice and one to Bob. In this paper, we have observed that the process of assigning identical particles to different users can be done in different ways.

If the operator of measure selective $M_m$ is local operation (LO), the whole quantum system has the same information to each of its parts, one located in Alice's lab and another in the Bob´s lab. Conversely, if the operator selective measurement $M_m$ is not a local operation, it increases entanglement between parts and the process is called entanglement by measurement (EBM).

In future work [13] we will discuss a more complete theory that contains both distinguishable and indistinguishable approaches to a system of identical particles, this will be done to satisfy a more complete theory in the context of the quantum information.

**References**
[1] Baym G 1969 *Lectures on Quantum Mechanics* (New York: W.A. Benjamin)
[2] Amico L, Fazo R, Osterloh A, and Vedral V 2008 *Rev. of Mod. Phys.* **80** 517
[3] Horodecki R, Horodecki P, Horodecki M, and Horodecki K 2009 *Rev. of Mod. Phys.* **81** 865
[4] Li Y S, Zeng B, Liu X S, and Long G L 2001 *Phys. Rev.* A **64** 0543022
[5] Schliemann J, Cirac J I, Kus M, Lewenstein M, and Loss D 2001 *Phys. Rev.* A **64** 0223035
[6] Shi Y 2003 *Phys. Rev.* A **67** 0243017
[7] Zanardi P 2002 *Phys. Rev. A* **65** 042101
[8] Zanardi P, and Wang X 2002 *J. Phys.* A **35** 7947
[9] Wiseman H M, and Vaccaro J A 2003 *Phys. Rev. Lett.* **91** 010404
[10] Vedral V 2003 *Cent. Eur. J. Phys.* **1** 197
[11] Cavalcanti D, França Santos M, Terra Cunha M O, Lunkes C, and Vedral V 2005 *Phys. Rev.* A **72** 062307
[12] Herbut F 2001 *Am. J. Phys.* **69** 207
[13] To be submitted.
[14] Nielsen M A, and 2000 *Quantum Computation and Quantum Information* (Cambridge: Cambridge University Press)
[15] Bose S, and Home D 2002 *Phys. Rev. Lett.* **88** 050401